\RequirePackage[switch]{lineno_babar_revtex}
\documentclass[twocolumn,showpacs,aps,prl,superscriptaddress,floatfix]{revtex4}
\usepackage{graphicx}
\usepackage{verbatim}

\usepackage{dcolumn}
\usepackage{amsmath}
\usepackage{epsfig}
\usepackage{subfigure}

\input babarsym.tex

\def\epem{e^+e^-}
\def\mpmm{\mu^+\mu^-}

\def\ma{m_a}
\def\MeV{\mev}
\def\GeV{\gev}

\def\figurebox#1#2#3{%
    \def\arg{#3}%
    \ifx\arg\empty
    {\hfill\vbox{\hsize#2\hrule\hbox to #2{\vrule\hfill\vbox to #1{\hsize#2\vfill}\vrule}\hrule}\hfill}%
    \else
    {\hfill\epsfbox{#3}\hfill}%
    \fi}

\begin{document}

\pagestyle{plain}

\begin{flushleft}
\babar-PUB-21/006\\
SLAC-PUB-17631
\end{flushleft}

\title{{\large \bf Search for an Axion-Like Particle in $B$ Meson Decays}}

\author{J.~P.~Lees}
\author{V.~Poireau}
\author{V.~Tisserand}
\affiliation{Laboratoire d'Annecy-le-Vieux de Physique des Particules (LAPP), Universit\'e de Savoie, CNRS/IN2P3,  F-74941 Annecy-Le-Vieux, France}
\author{E.~Grauges}
\affiliation{Universitat de Barcelona, Facultat de Fisica, Departament ECM, E-08028 Barcelona, Spain }
\author{A.~Palano}
\affiliation{INFN Sezione di Bari, I-70126 Bari, Italy}
\author{G.~Eigen}
\affiliation{University of Bergen, Institute of Physics, N-5007 Bergen, Norway }
\author{D.~N.~Brown}
\author{Yu.~G.~Kolomensky}
\affiliation{Lawrence Berkeley National Laboratory and University of California, Berkeley, California 94720, USA }
\author{M.~Fritsch}
\author{H.~Koch}
\author{T.~Schroeder}
\affiliation{Ruhr Universit\"at Bochum, Institut f\"ur Experimentalphysik 1, D-44780 Bochum, Germany }
% ...
\collaboration{The \babar\ Collaboration}
\noaffiliation

\begin{abstract}
Axion-like particles (ALPs)  are predicted in many extensions of the Standard Model, and their masses can 
naturally be well below the electroweak scale. In the presence of couplings to electroweak bosons, these 
particles could be emitted in flavor-changing $B$ meson decays. We report herein a search for an ALP, 
$a$, in the reaction $B^\pm\rightarrow K^\pm a$, $a\rightarrow\gamma\gamma$ using data collected by the 
\babar\ experiment at SLAC. No significant signal is observed, and 90\% confidence level upper limits on the 
ALP coupling to electroweak bosons are derived as a function of ALP mass, improving current constraints 
 by several orders of magnitude 
in the range $0.175\GeV < m_a < 4.78\GeV$.
\end{abstract}


\maketitle

\setcounter{footnote}{0}

The physics of spontaneous symmetry breaking drives much of the phenomenology of the Standard Model (SM). For 
instance, the Higgs mechanism gives mass to the fermions and weak gauge bosons of the SM, while the spontaneous 
breaking of approximate chiral global symmetries gives rise to pseudo-Goldstone bosons, such as the pions. Many extensions of the 
SM feature anomalous global symmetries whose spontaneous breaking leads to new pseudo-Goldstone bosons 
known as axion-like particles (ALPs)~\cite{Peccei:1977hh,Peccei:1977ur,Weinberg:1977ma,Wilczek:1977pj}. Such 
particles are ubiquitous in beyond-the-SM theories, such as supersymmetry~\cite{Frere:1983ag,Nelson:1993nf,Bagger:1994hh}, 
as well as in string theory~\cite{Witten:1984dg,Conlon:2006tq,Svrcek:2006yi,Arvanitaki:2009fg}. Potentially, ALPs could  resolve several
outstanding issues related to the naturalness of SM parameters,
 such as the strong $CP$ problem~\cite{Peccei:1977hh,Peccei:1977ur,Weinberg:1977ma,Wilczek:1977pj} 
or the hierarchy problem~\cite{Graham:2015cka}, and they may also serve as mediators 
to dark sectors~\cite{Nomura:2008ru,Freytsis:2010ne,Dolan:2014ska,Hochberg:2018rjs}. Consequently, ALPs have motivated a 
large number of searches in experimental particle physics and cosmology~\cite{Essig:2013lka,Marsh:2015xka,Graham:2015ouw,Irastorza:2018dyq}. 

In the simplest models, ALPs predominantly couple  to pairs of SM gauge bosons. While the photon and gluon 
couplings are already significantly constrained by collider and beam-dump experiments for ALP masses in the MeV--GeV 
range~\cite{units,Bergsma:1985qz,Riordan:1987aw,Bjorken:1988as,Blumlein:1990ay,Abbiendi:2002je,Aloni:2018vki,BelleII:2020fag}, the coupling to $W^\pm$ 
bosons is less explored. This coupling leads to ALP production in flavor-changing neutral-current decays, which can serve as powerful discovery modes. For example, flavor-changing $B$ meson and kaon decays already provide the most stringent 
bounds on invisibly decaying ALPs over a range of masses~\cite{Izaguirre:2016dfi}. The search presented here is the first for visibly decaying ALPs produced in $B$ meson decays. Its sensitivity complements existing studies of $K\rightarrow \pi\gamma\gamma$ \cite{Artamonov:2005ru,Abouzaid:2008xm,Ceccucci:2014oza}, which have been conservatively reinterpreted to obtain limits on ALP couplings~\cite{Izaguirre:2016dfi}.

In the following, we  consider a minimal ALP ($a$) model with coupling $g_{aW}$ to the $\mathrm{SU}(2)_{W}$ 
gauge-boson field strengths, $W_{\mu\nu}^b$, and Lagrangian
\begin{linenomath}
\begin{equation}\label{eq:alp_lagrangian}
\mathcal{L} = -\frac{g_{aW}}{4}\,a\,W^b_{\mu\nu} \tilde{W}^{b\mu\nu},
\end{equation}
\end{linenomath}
where $\tilde{W}^{b\mu\nu}$ is the dual field-strength tensor. This coupling leads to the production of ALPs at one loop in the 
process $B^\pm\rightarrow K^\pm a$, where the ALP is emitted from an internal $W^\pm$ boson~\cite{Izaguirre:2016dfi}. Electroweak symmetry breaking and the resulting gauge-boson 
mixing  generates an ALP coupling to a pair of photons, and the branching fraction for $a\rightarrow\gamma\gamma$ in this model is nearly 100\% for $\ma < m_W$. The same ALP production and decay modes also occur in models with  axion couplings to gluons \cite{Bertholet:2021hjl}.

We report herein the first search for an ALP in the reaction $B^\pm\rightarrow K^\pm a$, $a\rightarrow\gamma\gamma$ in the range 
$0.175 \GeV < m_a < m_{B^+} -m_{K^+} \approx 4.78 \GeV$, excluding the mass intervals 0.45--$0.63 \GeV$ and 
0.91--$1.01 \GeV$ because of large peaking backgrounds from $\eta$ and $\eta'$ mesons, respectively. Note that existing searches already constrain $m_a<0.1\GeV$ in the range of couplings to which our search is sensitive \cite{Bergsma:1985qz,Riordan:1987aw,Bjorken:1988as,Blumlein:1990ay,Abbiendi:2002je}, while the mass range $0.1\GeV<m_a<0.175\GeV$ is excluded from our analysis due to large peaking $\pi^0$ contributions. The $B^\pm\rightarrow K^\pm a$, $a\rightarrow\gamma\gamma$ product branching fraction is 
measured assuming all signal observed is produced in $B^\pm\rightarrow K^\pm a$ with $a$ decaying promptly. However, the ALP has a decay width  $\Gamma_a = g_{aW}^2 m_a^3 \sin^4\theta_{\rm W}/64\pi$, where $\theta_{\rm W}$ is the weak mixing angle, and the present search has sensitivity to couplings predicting long-lived ALPs for $\ma<2.5\GeV$.   We  
additionally determine the branching fraction for $c\tau_a$ values of 1, 10, and $100 \mm$ in this mass range. 

The search is 
based on a sample of $4.72\times10^8$ $B\bar{B}$ meson pairs corresponding to 424 fb$^{-1}$ of integrated luminosity collected at 
the $\Y4S$ resonance by the \babar\ detector at the PEP-II $\epem$ storage ring at the SLAC National Accelerator Laboratory~\cite{Lees:2013rw}. The \babar\ detector is 
described in detail elsewhere~\cite{Bib:Babar,TheBABAR:2013jta}. A small sample,  corresponding to  8\% of the total data set, 
is used to optimize the search strategy and is subsequently discarded.

Signal Monte Carlo (MC) events  are simulated using \textsc{EvtGen}~\cite{Lange:2001uf}, with MC samples generated at 24 masses (from $0.1-4.8\GeV$)
for promptly decaying ALPs and 16 masses for long-lived ALPs (from $0.1-2.5\GeV$). 
We simulate the following reactions to study the background: 
$\epem \rightarrow \epem (\gamma)$ (\textsc{BHWIDE}~\cite{Jadach:1995nk}), $\epem \rightarrow \mpmm (\gamma)$, 
$\epem \rightarrow \tau^+ \tau^- (\gamma)$ (\textsc{KK} with \textsc{TAUOLA} library~\cite{Jadach:2000ir,Jadach:1993hs}), continuum 
$\epem \rightarrow q\overline{q}$ with $q=u,d,s,c$ (\textsc{JETSET}~\cite{Sjostrand:1993yb}), and 
$\epem \rightarrow B\bar{B}$ (\textsc{EvtGen}). Each background MC sample is weighted to match the luminosity of the data set. The detector acceptance and reconstruction efficiencies are estimated with a simulation 
based on \textsc{GEANT4}~\cite{Agostinelli:2002hh}.

We reconstruct signal $B^\pm$ candidates by combining a pair of photons with a track identified as a kaon by particle 
identification algorithms~\cite{Bib:Babar}. All other reconstructed tracks and neutral clusters in the event are collectively 
referred to as the rest of the event (ROE). To suppress backgrounds, we require an energy-substituted mass 
$m_{\rm ES}= \sqrt{(s/2+\vec{p}_i\cdot\vec{p}_B)^2/E_i^2-p^{2}_B} > 5.0 \GeV$ and an energy difference $\DeltaE = |\sqrt{s}/2-E^{*}_B| < 0.3 \GeV$, where 
$\sqrt{s}$ denotes the center-of-mass (CM) energy, $\vec{p}_B$ and $E_B$ are the $B^\pm$ momentum and energy in the lab frame, $E_B^*$ is the $B^\pm$ energy in the CM frame, and $E_i$ and $\vec{p}_i$ are the energy and momentum of the initial state in the lab frame. A kinematic fit is  performed on the selected $B^\pm$ candidates, requiring the photon and kaon 
candidates to originate from the measured beam interaction region, and constraining their total energy and invariant mass to the 
nominal $B^\pm$ meson mass and measured CM beam energy.

Two boosted decision trees (BDTs)~\cite{BDT} are used to further
  separate signal from each of the main backgrounds:~one BDT is trained using continuum MC background events and the other
 using $B^+B^-$ MC background events. For the signal sample, we combine events from all simulated ALP masses with prompt decays to obtain a uniform distribution in diphoton invariant mass $(m_{\gamma\gamma})$. Each BDT includes the following 13  observables:~invariant mass of the ROE; 
 cosine of the 
angle between two sphericity axes, one computed with the $B^\pm$ constituents and the other with the ROE; second Legendre moment of the ROE, calculated relative to the $B^\pm$ thrust axis; 
$m_{\rm ES}$ and $\DeltaE$;  particle 
identification information for the $K^\pm$;  helicity   angle of the $K^\pm$, which is the angle between the $K^\pm$ and the $\Upsilon(4S)$ as measured in the $B^\pm$  frame;
helicity  angle and energy of the most energetic photon forming the $a$;  three invariant masses $m(\gamma_i\gamma_j^{P})$, where $\gamma_i$ is an ALP-daughter photon, $\gamma_j^{P}$ is a photon in the ROE, and  $\gamma_i$ and $\gamma_j^{P}$ are chosen so that  $m(\gamma_i\gamma_j^{\rm P})$ is closest to the nominal mass of each of $P=\pi^0,\eta,\eta'$; and, multiplicity of neutral candidates in the event.

\begin{figure}[t]
\begin{center}
  \includegraphics[width=0.48\textwidth]{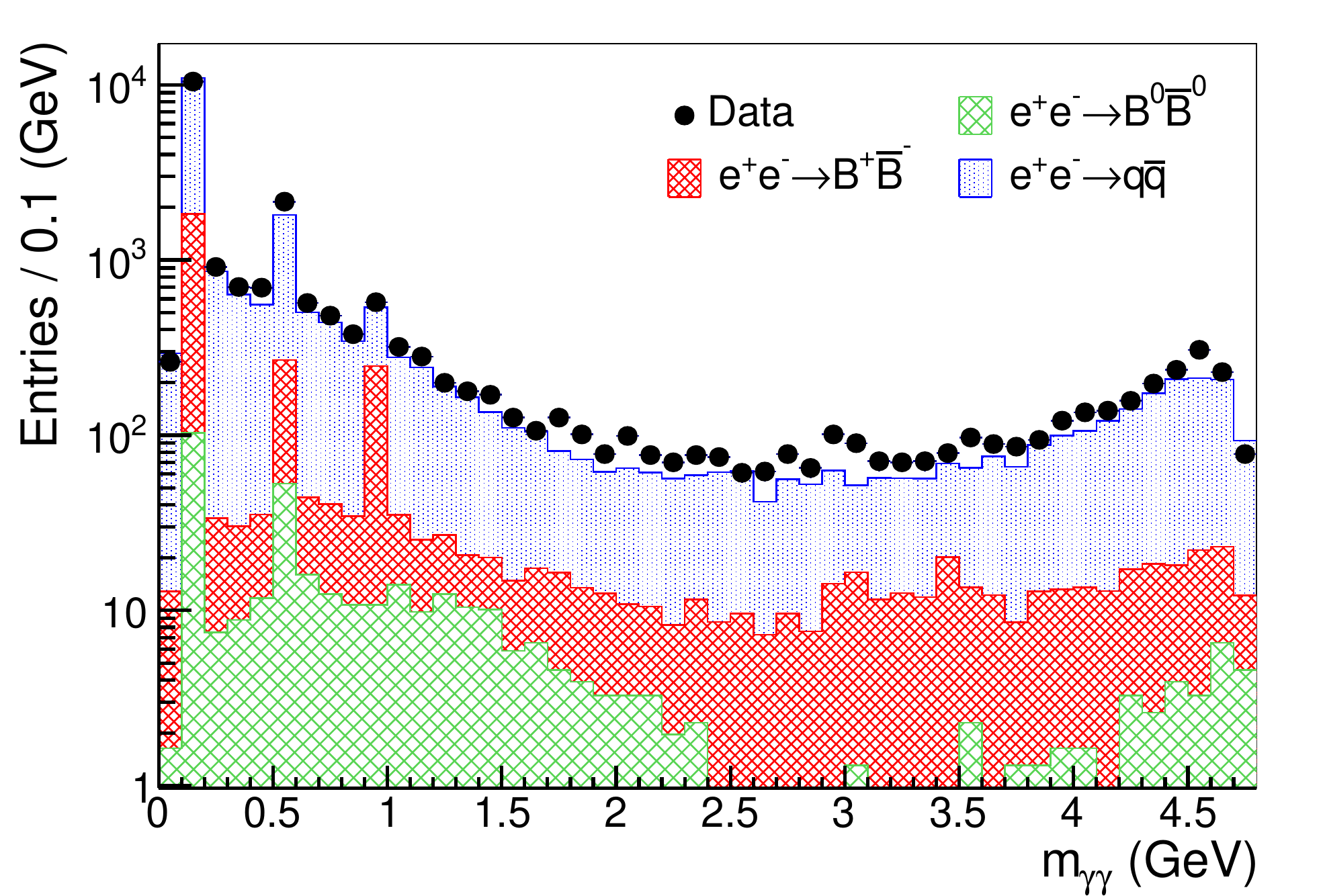}
\end{center}
\caption
{The diphoton mass distribution of ALP candidates, together with Monte Carlo predictions of leading 
background processes normalized to the data luminosity.}
\label{Fig1}
\end{figure}

The BDT score distributions for data, signal MC, and background MC are provided in Ref.~\cite{SPM}.  For our final signal region selection, we apply the criteria on the two BDT scores shown in Ref.~\cite{SPM}, allowing multiple candidates per event.  The BDT selection criteria are independent of the  ALP mass hypothesis. The signal efficiency estimated from MC varies between $2\%$ for $m_a=4.78\GeV$ to $33\%$ for $m_a = 0.3\gev$.  The resulting $m_{\gamma\gamma}$ distribution is shown in Fig.~\ref{Fig1}. 

The background is dominated by continuum 
events and by peaking contributions from $B^\pm\rightarrow K^\pm h^0$ and $B^\pm\rightarrow \pi^\pm h^0$ 
decays where $h^0=\pi^0,\eta,\eta'$. Non-resonant $B^\pm\rightarrow K^\pm\gamma\gamma$ decays and $B^\pm\rightarrow K^*\gamma,\,K^*\rightarrow K\gamma$ decays are negligible, as they have total branching fractions $\lesssim10^{-7}$ \cite{Reina:1997my,pdg} and do not give a  peak in $m_{\gamma\gamma}$. The $B^\pm\rightarrow K^\pm\eta_c,\,\eta_c\rightarrow\gamma\gamma$ decay is not included in our background MC; we observe an excess at the $\eta_c$ mass, with a local significance of $2.6 \sigma$ as determined by the signal extraction procedure defined below. The measured product branching fraction is consistent with the  world average value of ${\cal B}(B^\pm\rightarrow K^\pm \eta_c){\cal B}(\eta_c\rightarrow\gamma\gamma)$~\cite{pdg}. Because of the relatively small $\eta_c$ background compared to the $\pi^0$, $\eta$, and $\eta'$, we do not exclude signal mass hypotheses in the vicinity of the $\eta_c$ mass.

We extract the signal yield of promptly decaying ALPs  by performing a series of unbinned maximum likelihood fits of a hypothetical signal peak over a smooth background to the data shown in Fig.~\ref{Fig1}. We perform fits for 461 signal mass hypotheses with a scan step size equal to the signal 
resolution, $\sigma_{\gamma\gamma}$. The latter is determined by fitting the signal sample at each simulated ALP mass with a double-sided Crystal Ball 
function~\cite{CrystalBall} and interpolating the results to the remaining mass hypotheses. The resolution ranges from 
$8 \MeV$ near $m_a=0.175\GeV$ to $14\MeV$ near $m_a = 2 \GeV$, and decreasing back to $2\MeV$ near $m_a = 4.78\GeV$ 
as a result of the constraint imposed on the mass of the $B^\pm$ meson candidate in the kinematic fit. The MC predictions are 
validated using a sample of $B^\pm \rightarrow K^\pm \pi^0$ and $B^\pm \rightarrow K^\pm \eta$ decays. The simulated $\pi^0$ 
and $\eta$ mass resolutions agree with the data to within 3\%.

Each unbinned likelihood fit is performed over an $m_{\gamma\gamma}$  interval with a width in the range (24--60)$\sigma_{\gamma\gamma}$. The mass-dependent interval width is chosen to be sufficiently broad as to  fix the continuum background shape. We have verified that our results are independent of minor variations of the fit interval widths.
The probability density function (pdf) includes contributions from signal, continuum background components, and, where needed, peaking components describing 
the $\pi^0$, $\eta$, $\eta'$, and $\eta_c$.

\begin{figure}[t]
\begin{center}
  \includegraphics[width=0.48\textwidth]{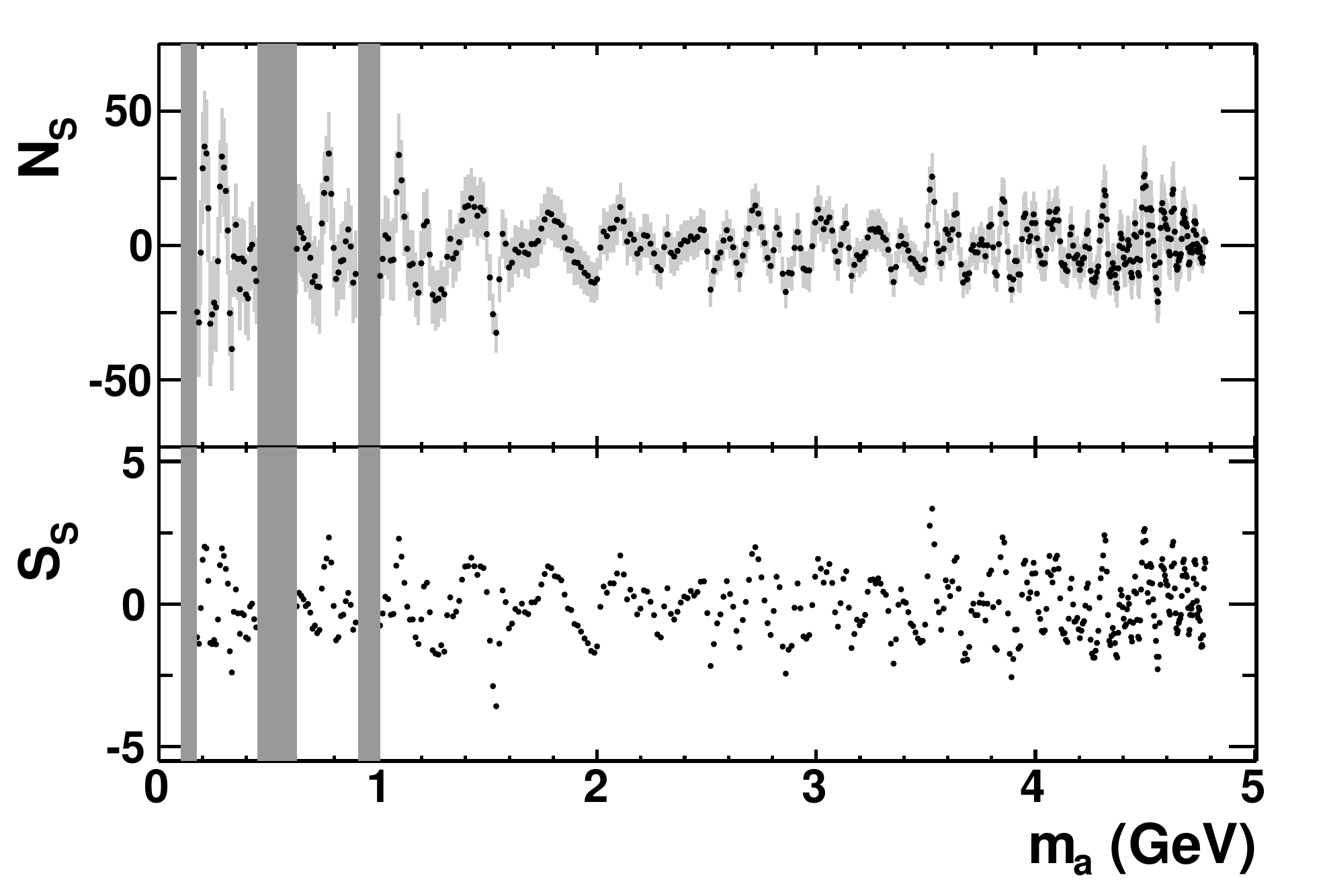}
\end{center}
\caption{The distribution of signal events ($N_{\rm s}$) and local signal significance ($S_{\rm s}$) from fits as a function 
of $m_a$ for prompt ALP decays. The vertical
gray bands indicate the regions excluded from the search in the vicinity of the $\pi^0$, $\eta$, and $\eta'$ masses.}
\label{Fig2}
\end{figure}

The signal pdf is described by a non-parametric kernel 
density function modeled  from  signal MC and extrapolated between simulated mass points~\cite{Read:1999kh}. 
The continuum background is modeled for $m_a<4\GeV$ by the sum of a template derived from background MC and a first-order polynomial, with the normalization determined from the fit. At higher masses, only the first-order polynomial is needed to model the background. The data-to-MC ratio is approximately constant over each fit interval, and the residual differences are accommodated by the  linear polynomial.  The 
shapes of the  $\pi^0$, $\eta$, and $\eta'$ resonances are also modeled from background MC, while the $\eta_c$ is modeled using the signal MC mass distribution with a width broadened to match the $\eta_c$ natural linewidth. For the $\pi^0$, $\eta$, and $\eta'$ background components, the normalization is determined from the fit to data, while the normalization of the $\eta_c$ component is fixed to the product of  the world-average value of ${\cal B}(B^\pm\rightarrow K^\pm \eta_c){\cal B}(\eta_c\rightarrow\gamma\gamma)$ and the signal efficiency evaluated at this mass. This allows us to measure an ALP signal rate for $m_a\approx m_{\eta_c}$ while simultaneously accounting for events from $B^\pm\rightarrow K^\pm \eta_c,\eta_c\rightarrow\gamma\gamma$  decays. We have verified that our signal extraction procedure is robust against changes in the background model by varying the order of the polynomial component of the continuum background.

To assess systematic uncertainties in the MC-derived continuum and peaking background components, we fit the relative normalizations of different background components (continuum $q\bar{q},\,B^+B^-,\,B^0\bar{B}^0$) to data rather than fixing each component's normalization to match the luminosity of the total data set, and we repeat our signal extraction procedure with the re-weighted MC-derived templates. We also propagate the uncertainties in the resolution of the peaking components and in the uncertainties in the world-average value of the $\eta_c$ linewidth. For the $\eta_c$  model, we  assess a systematic uncertainty originating from uncertainties in  ${\cal B}(B^\pm\rightarrow K^\pm \eta_c){\cal B}(\eta_c\rightarrow\gamma\gamma)$ by varying the $\eta_c$ normalization within the uncertainties in the world-average value.  The systematic uncertainty in the 
signal yield resulting from variations in the continuum (peaking) background shape due to re-fitting the component normalizations is estimated to be, on average, 1\% (2\%) of the corresponding statistical uncertainty. 

We further assess systematic uncertainties associated with our signal model.  The 
systematic uncertainty in the signal yield derived from our extrapolation procedure between simulated mass points is estimated to be, on average, 4\% of the corresponding statistical uncertainty.   We assess a signal resolution systematic uncertainty by repeating our fits with a signal shape whose width is varied within the mass resolution uncertainty, leading to a  signal resolution systematic uncertainty that is, on average, 3\% of the statistical uncertainty. A 6\% relative systematic 
uncertainty in the signal efficiency is derived from the data-to-MC ratio for events in the vicinity of the $\eta'$ resonance. 

The fitted signal yields and statistical significances are shown in Fig.~\ref{Fig2}. The largest local 
significance of $3.3\sigma$ is observed near $m_a = 3.53 \GeV$ with a global significance of $1.1\sigma$ 
after including trial factors \cite{Gross:2010qma}, consistent with the null hypothesis. Background-only fits to the $m_{\gamma\gamma}$
spectrum are shown over the whole mass range in Ref.~\cite{SPM}.

To further validate the signal extraction procedure, we measure the $B^\pm\rightarrow K^\pm h^0, h^0\rightarrow \gamma \gamma$ 
($h^0=\pi^0,\eta,\eta',\eta_c$) product branching fractions by treating the peaks as signal, extracting the number of events in the peak using the fitting procedure described above, and subtracting non-peaking background whose magnitude is determined from MC.  The results are found to be compatible with the current world averages~\cite{pdg} within uncertainties.

\begin{figure}[t]
\begin{center}
  \includegraphics[width=0.48\textwidth]{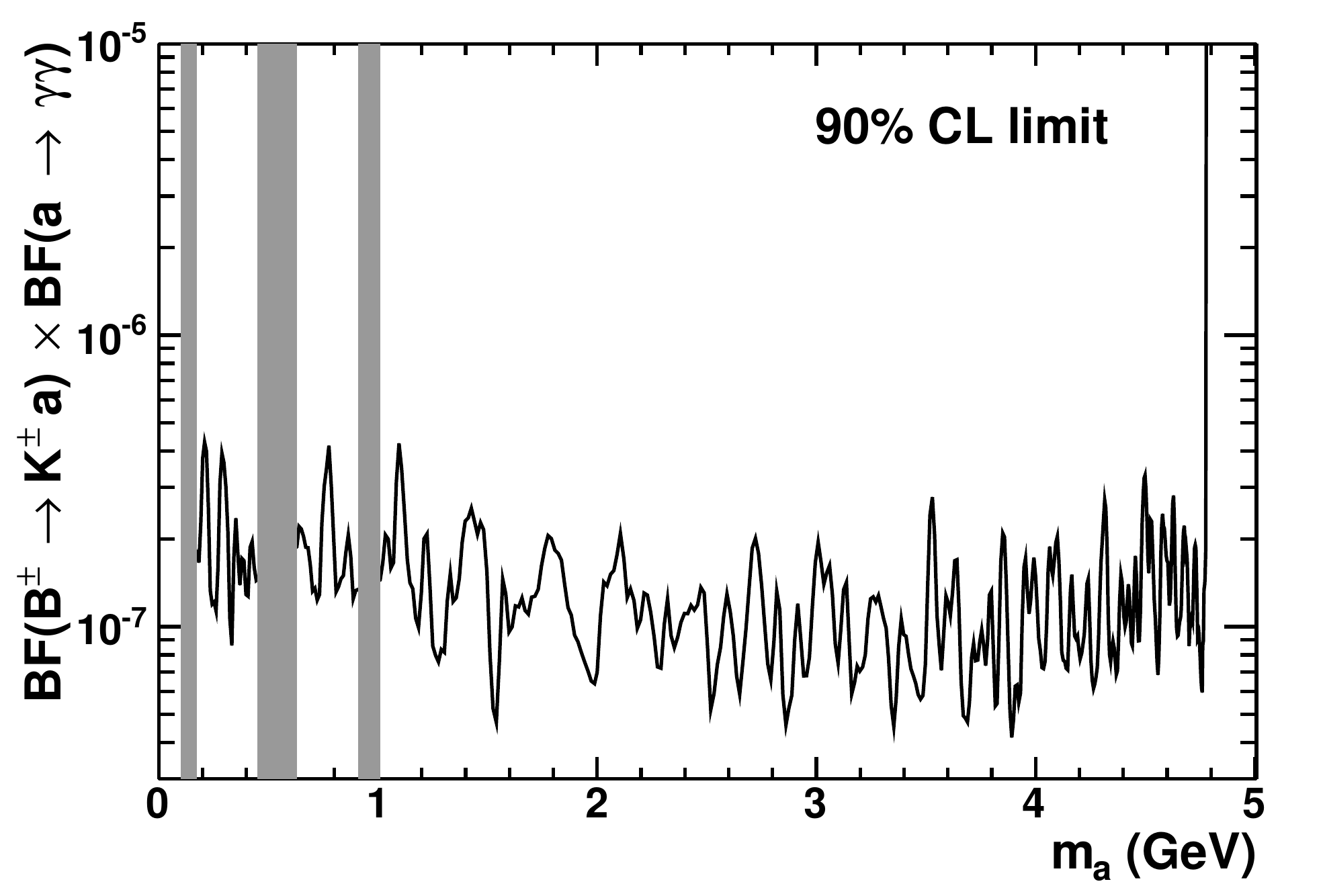}
\end{center}
\caption{90\% CL upper limits on the $B^\pm\rightarrow K^\pm a$ branching fraction assuming promptly decaying ALPs. The 
vertical gray bands indicate the regions excluded from the search in the vicinity of the $\pi^0$, $\eta$, and $\eta'$ masses.}
\label{Fig3}
\end{figure}

In the absence of significant signal, Bayesian upper limits at 90\% confidence level (CL) on 
${\cal B}(B^\pm\rightarrow K^\pm a){\cal B}(a\rightarrow\gamma\gamma)$ are derived with a uniform positive prior in the product branching fraction.
We have verified that the limits are robust with respect to the choice of prior.
The systematic uncertainty is included in the limit calculation by convolving the likelihood function with a Gaussian having a width  equal to the 
systematic uncertainty. Uncertainties in the luminosity (0.6\%)~\cite{Lees:2013rw} and from the limited statistical
  precision of simulated samples (1\%) are included as well. The resulting limits on the branching fraction product assuming promptly decaying ALPs are 
displayed in Fig.~\ref{Fig3}.

Our search targets promptly decaying ALPs. However,  ALPs can be long lived at small masses and coupling, and we assess how our signal efficiency and resolution are affected for ALP proper decay lengths of $c\tau_a=1$, 10, and 100 mm. These decay lengths range from nearly prompt decays for which the efficiency and resolution are comparable to the zero-lifetime signal, through to the longest values to which our analysis is sensitive. We  measure the $B^\pm\rightarrow K^\pm a$ branching fraction for each decay length. We restrict 
this study to the mass range for which we obtain sensitivity to couplings that give rise to long-lived ALPs, namely $m_a<2.5\GeV$.  Long-lived ALPs induce a non-negligible bias in the measurement of $m_{\gamma\gamma}$, and the resolution is significantly impacted, ranging from $15\MeV$ near $m_a=0.175\GeV$ to $28\MeV$ near $2\GeV$ 
for $c\tau_a=100\mm$. For $c\tau_a=100\mm$, we only consider mass hypotheses $m_a\ge0.2\GeV$,  because there is a significant overlap between the signal mass distribution and the $\pi^0$ background for lower ALP masses.

\begin{figure}[t]
\begin{center}
  \includegraphics[width=0.48\textwidth]{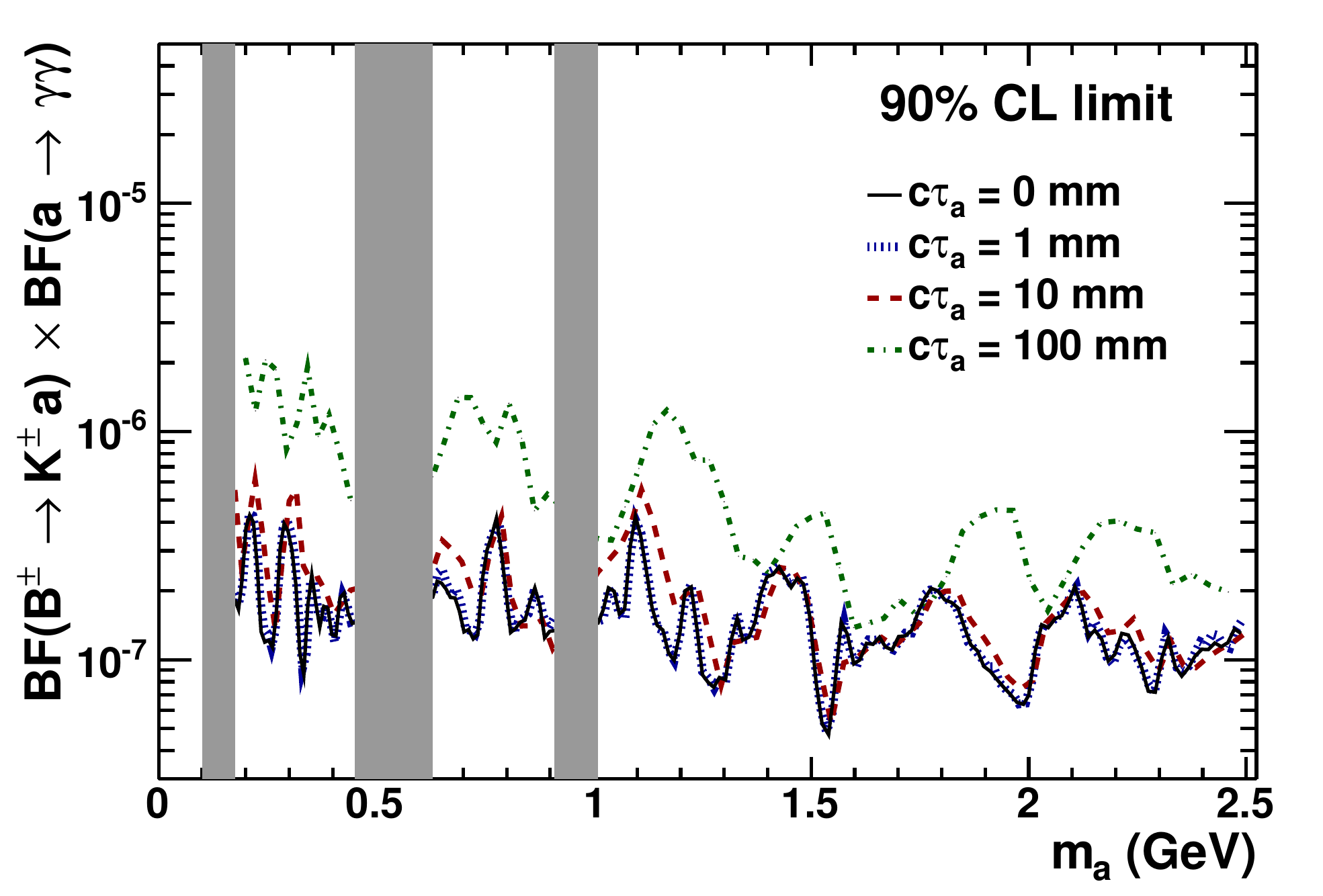}
\end{center}
\caption{The 90\% CL upper limits on the $B^\pm\rightarrow K^\pm a$ branching fraction for $m_a<2.5\GeV$ and  $c\tau_a$ between 0 and 100 mm. The vertical gray bands indicate the regions excluded from the search in the vicinity of the $\pi^0$, $\eta$, and $\eta'$ masses.}
\label{Fig5}
\end{figure}

The signal is extracted in the same manner as for the promptly decaying ALP, and the fitted signal yields and local statistical significances are shown in Ref.~\cite{SPM}.  The largest local significance is found to be  at $m_a=1.10\GeV$ and $c\tau_a=10 \mm$, with a global significance of less than one standard deviation.  Systematic uncertainties are assessed in the same manner as for the prompt analysis. The systematic uncertainty in the 
signal yield resulting from variations in the continuum (peaking) background shape due to re-fitting the component normalizations is larger for long-lived ALPs because of the long tail induced by the bias in the measurement of the signal $m_{\gamma\gamma}$ distribution, and is estimated to be, on average, 16\% (24\%) of the corresponding statistical uncertainty for $c\tau_a=100\mm$. The other systematic uncertainties are comparable in magnitude to the values for  prompt ALPs, and the total systematic uncertainty is  subdominant to the statistical uncertainty for all signal mass hypotheses.

The 90\% CL upper limits on ${\cal B}(B^\pm\rightarrow K^\pm a){\cal B}(a\rightarrow \gamma\gamma)$ 
are plotted in Fig.~\ref{Fig5}. The limits degrade at $c\tau_a=100\mm$ because of the broadening of the signal 
shape and lower efficiency. The $c\tau_a$ dependence of the limit is less pronounced at higher masses because the ALP is less 
boosted, leading to a shorter decay length in the detector. We use an interpolating function to obtain product branching fraction
limits for intermediate lifetimes between those shown in Fig.~\ref{Fig5}.

\begin{figure}[t]
\begin{center}
  \includegraphics[width=0.48\textwidth]{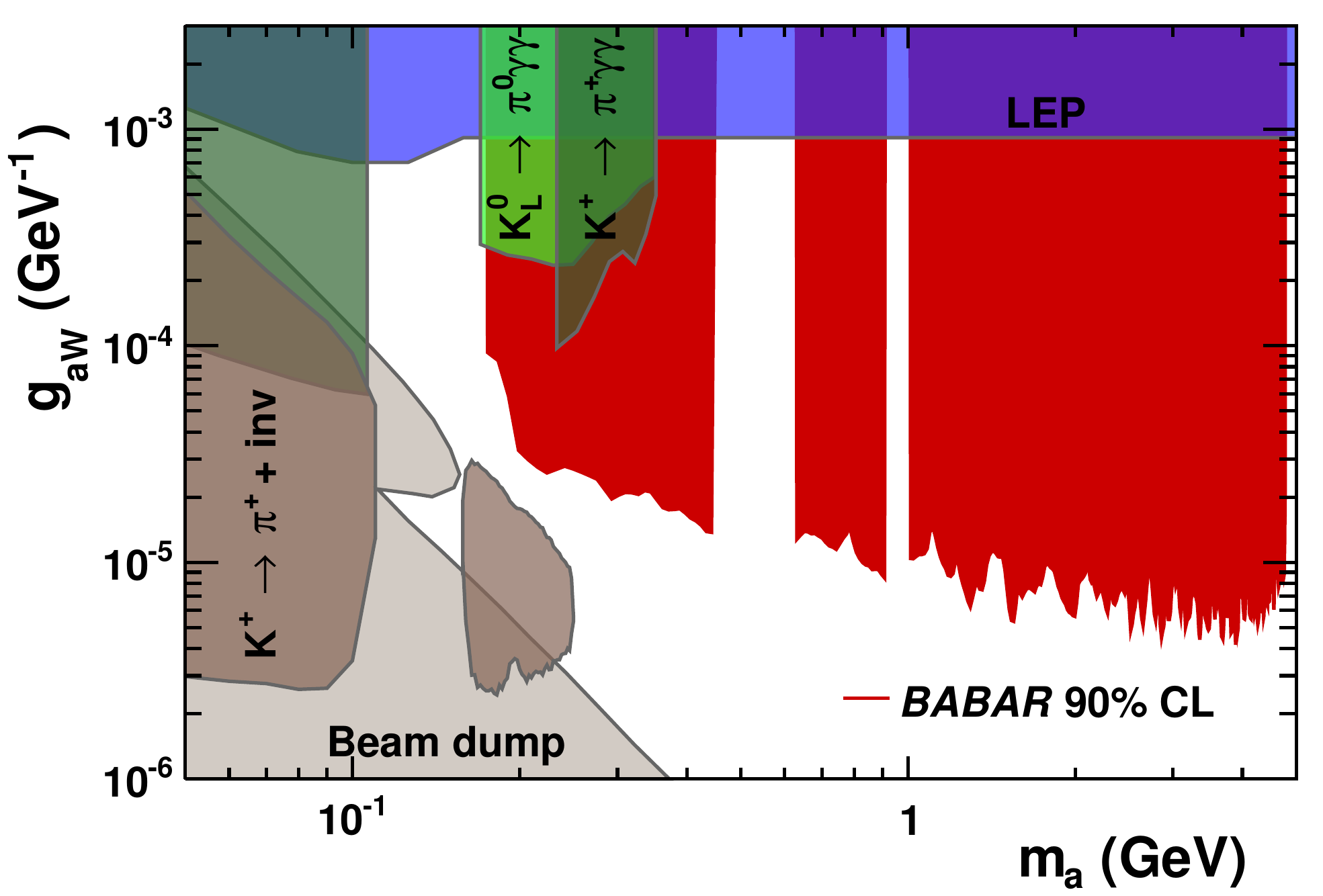}
\end{center}
\caption
{The 90\% CL upper limits on the coupling $g_{aW}$ as a function of the ALP mass (red), together with existing 
constraints \cite{Izaguirre:2016dfi} (blue, green, brown, and grey). }
\label{Fig6}
\end{figure}

The 90\% CL limits on the ALP coupling $g_{aW}$ are presented in Fig.~\ref{Fig6}. For each ALP mass hypothesis, we determine the value of $g_{aW}$ such that the 
calculated branching fraction is equal to the 90\%-CL-excluded branching fraction for the  lifetime predicted using the same value of $g_{aW}$. This is the excluded value of $g_{aW}$ shown in Fig.~\ref{Fig6}. The  90\% CL 
bounds on $g_{aW}$ extend below $10^{-5}\GeV^{-1}$ for many ALP masses, improving current constraints by more than two orders of magnitude. The strongest limit on the coupling at $m_a=0.2\GeV$ corresponds to a  lifetime of $c\tau_a=100\mm$, decreasing to $c\tau_a=1 \mm$ at $m_a=2.5\GeV$. Along with our limit, we show in Fig.~\ref{Fig6} existing constraints derived in Ref.~\cite{Izaguirre:2016dfi} from LEP, beam dump, and $K\rightarrow\pi\gamma\gamma$ searches. We have also re-interpreted a search for $K^\pm\rightarrow \pi^\pm X$ with invisible $X$ \cite{NA62:2020xlg}, which applies to our model if the ALP is sufficiently long lived that it  decays outside of the detector.
 
In summary, we report the first search for axion-like particles in the process $B^\pm\rightarrow K^\pm a, a\rightarrow\gamma\gamma$. 
The results strongly constrain ALP couplings to electroweak gauge bosons, improving upon current bounds by several orders of 
magnitude, except in the vicinity of the $\pi^0$, $\eta$, and $\eta'$  resonances. Our results demonstrate the sensitivity of flavor-changing neutral 
current probes of ALP production, which complement existing searches for the  ALP coupling to photons below the $B$ meson mass.

We are grateful for the 
extraordinary contributions of our \pep2\ colleagues in
achieving the excellent luminosity and machine conditions
that have made this work possible.
The success of this project also relies critically on the 
expertise and dedication of the computing organizations that 
support \babar.
The collaborating institutions wish to thank 
SLAC for its support and the kind hospitality extended to them. 
This work is supported by the
US Department of Energy
and National Science Foundation, the
Natural Sciences and Engineering Research Council (Canada),
the Commissariat \`a l'Energie Atomique and
Institut National de Physique Nucl\'eaire et de Physique des Particules
(France), the
Bundesministerium f\"ur Bildung und Forschung and
Deutsche Forschungsgemeinschaft
(Germany), the
Istituto Nazionale di Fisica Nucleare (Italy),
the Foundation for Fundamental Research on Matter (The Netherlands),
the Research Council of Norway, the
Ministry of Education and Science of the Russian Federation, 
Ministerio de Econom\'{\i}a y Competitividad (Spain), the
Science and Technology Facilities Council (United Kingdom),
and the Binational Science Foundation (U.S.-Israel).
Individuals have received support from 
the Marie-Curie IEF program (European Union) and the A. P. Sloan Foundation (USA). 


\clearpage
\newpage
\onecolumngrid
\begin{center} {\bf EPAPS Material} \end{center}
The following includes supplementary material for the Electronic Physics Auxiliary Publication Service.

\begin{figure}[h]
\begin{center}
  \includegraphics[width=0.9\textwidth]{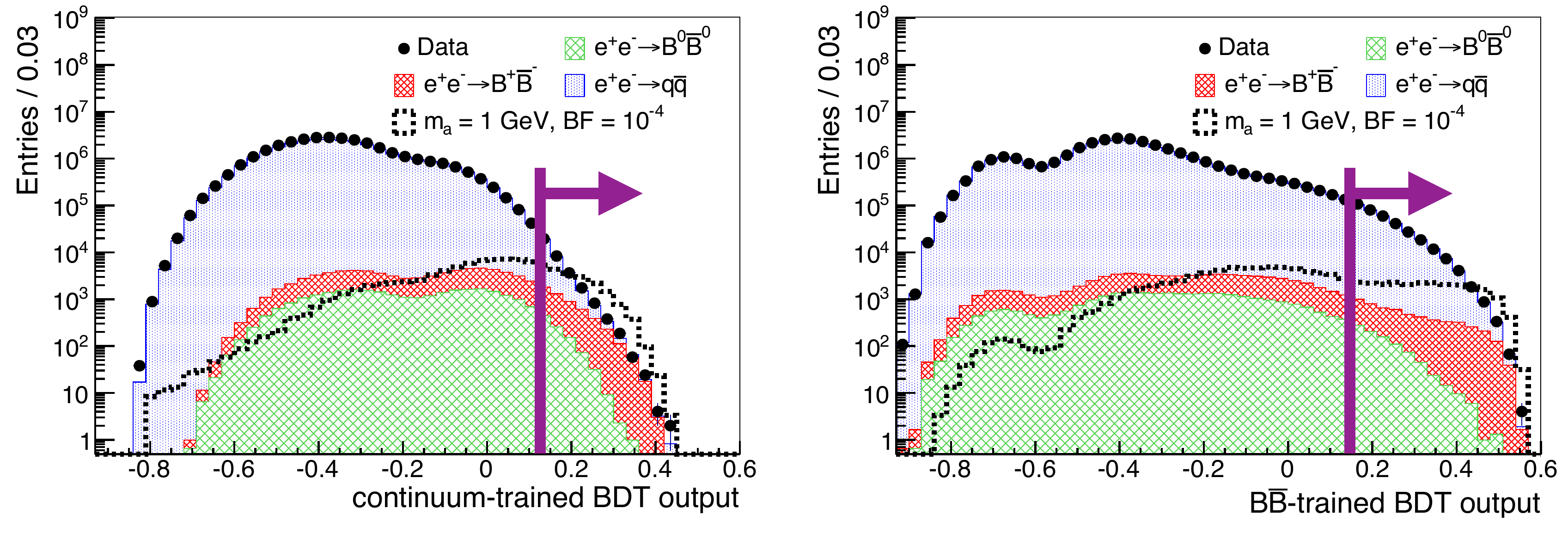}
\end{center}
\caption{Boosted decision tree classifier scores for ALP candidates in data, signal MC with $m_a=1\GeV$ normalized to a branching fraction of $10^{-4}$, and background MC normalized to the data luminosity. The classifier scores are for BDTs trained on (left) continuum MC; (right) $B^+B^-$ MC.  The
selections used in the analysis are indicated for each distribution by the arrow.}
\label{epaps1}
\end{figure}

      \clearpage
      \newpage

\begin{figure}[h]
\begin{center}
  \includegraphics[width=0.3\textwidth]{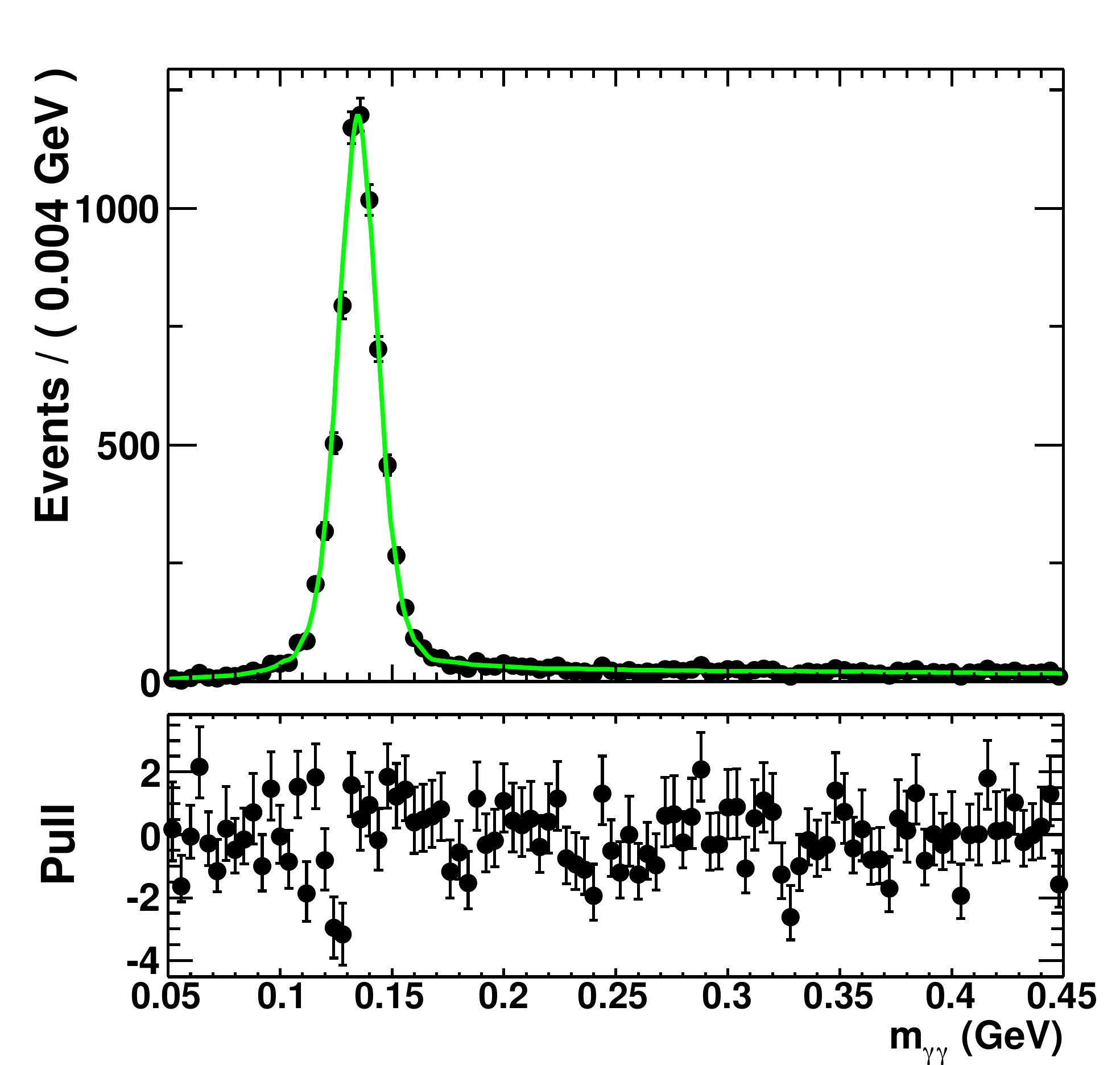}
    \includegraphics[width=0.3\textwidth]{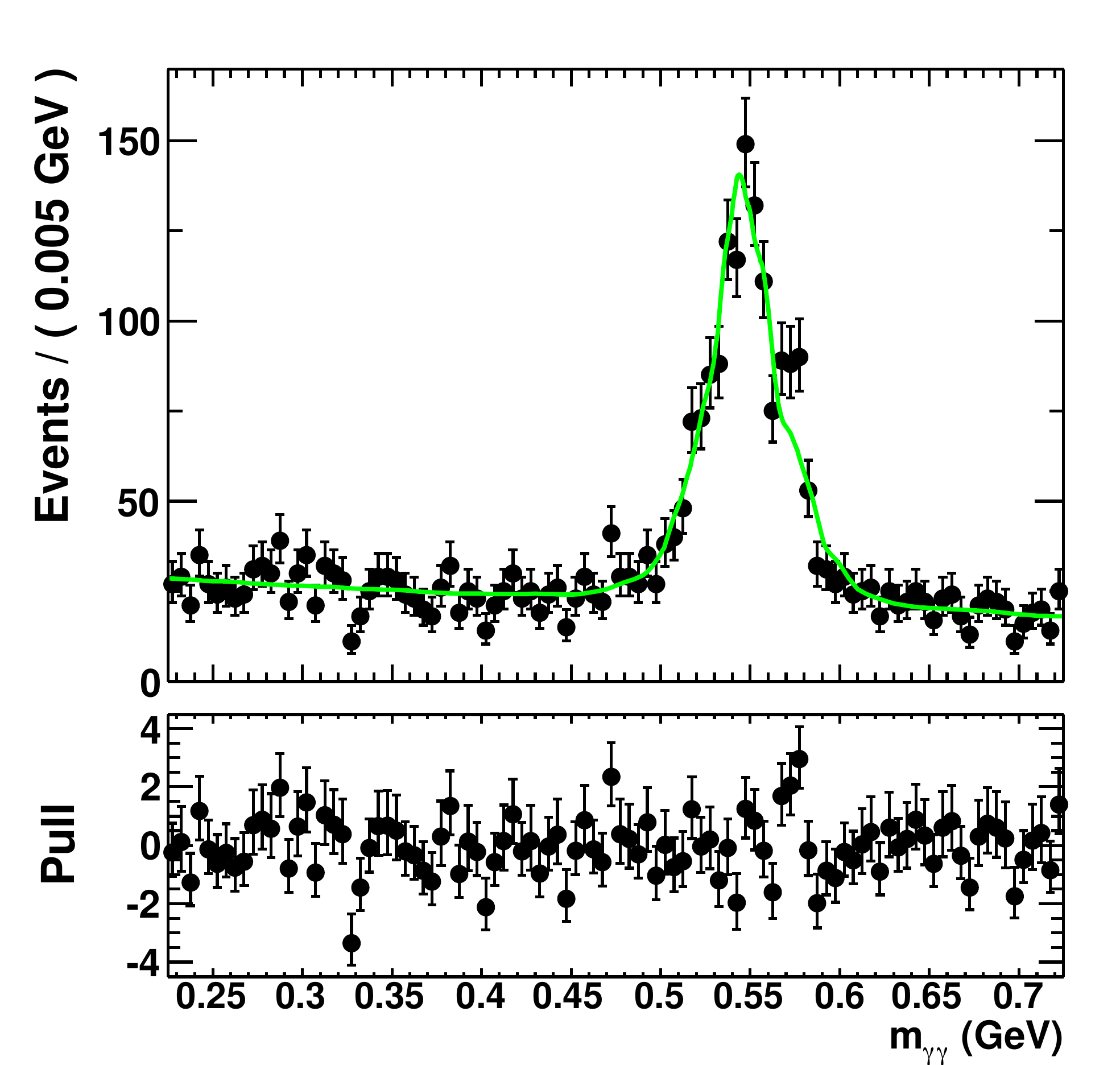}
      \includegraphics[width=0.3\textwidth]{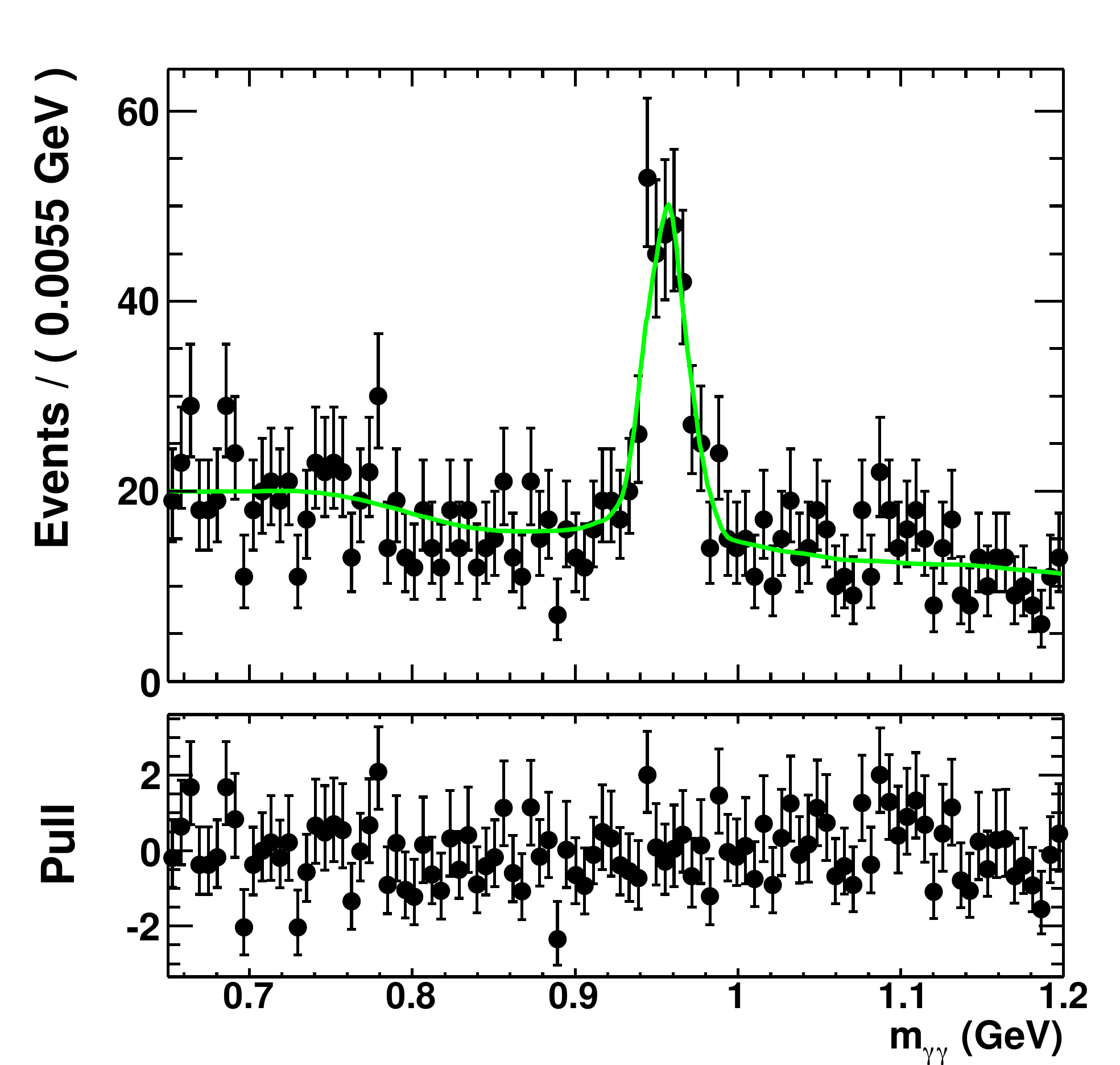}\\
        \includegraphics[width=0.3\textwidth]{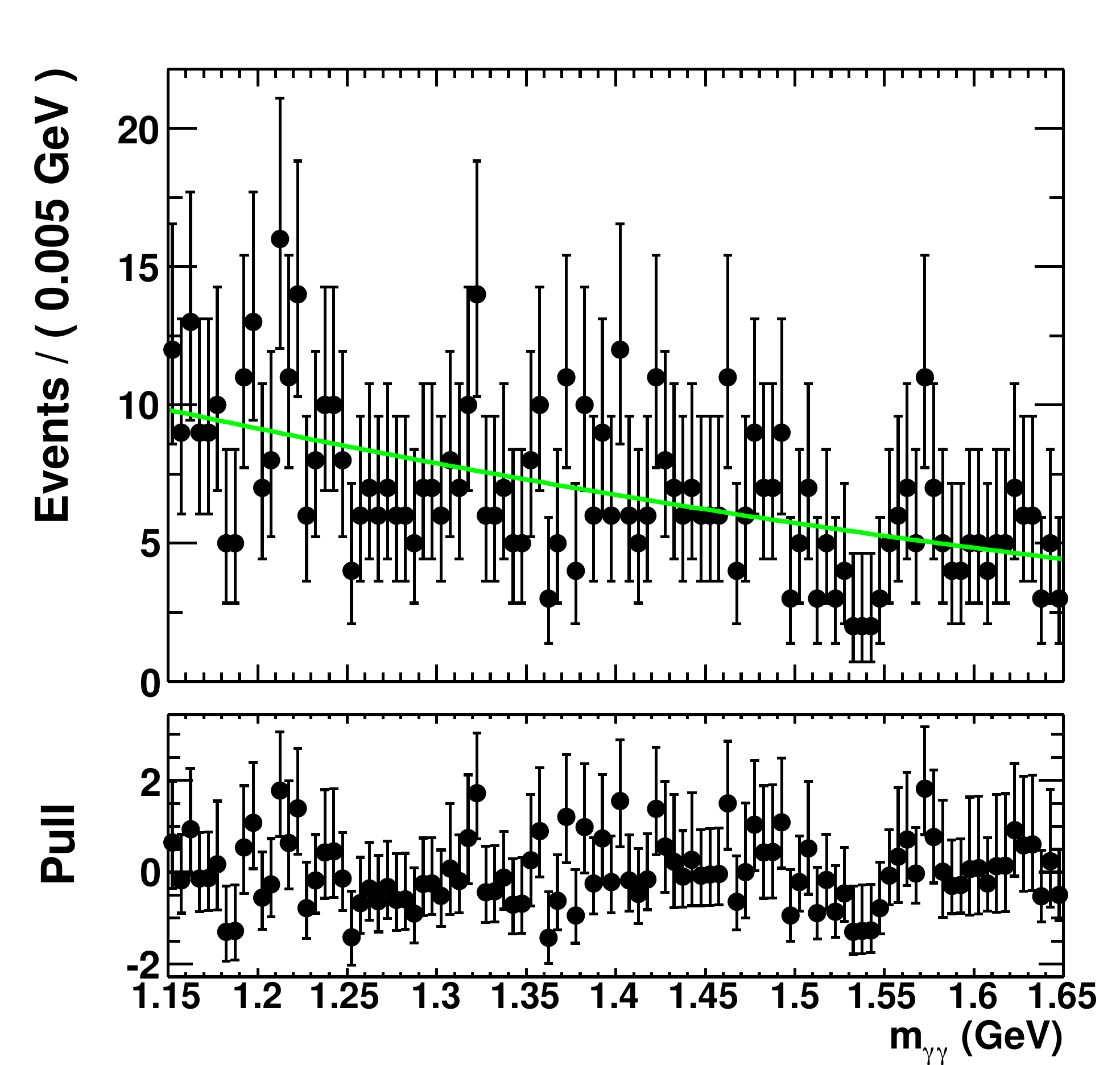}
    \includegraphics[width=0.3\textwidth]{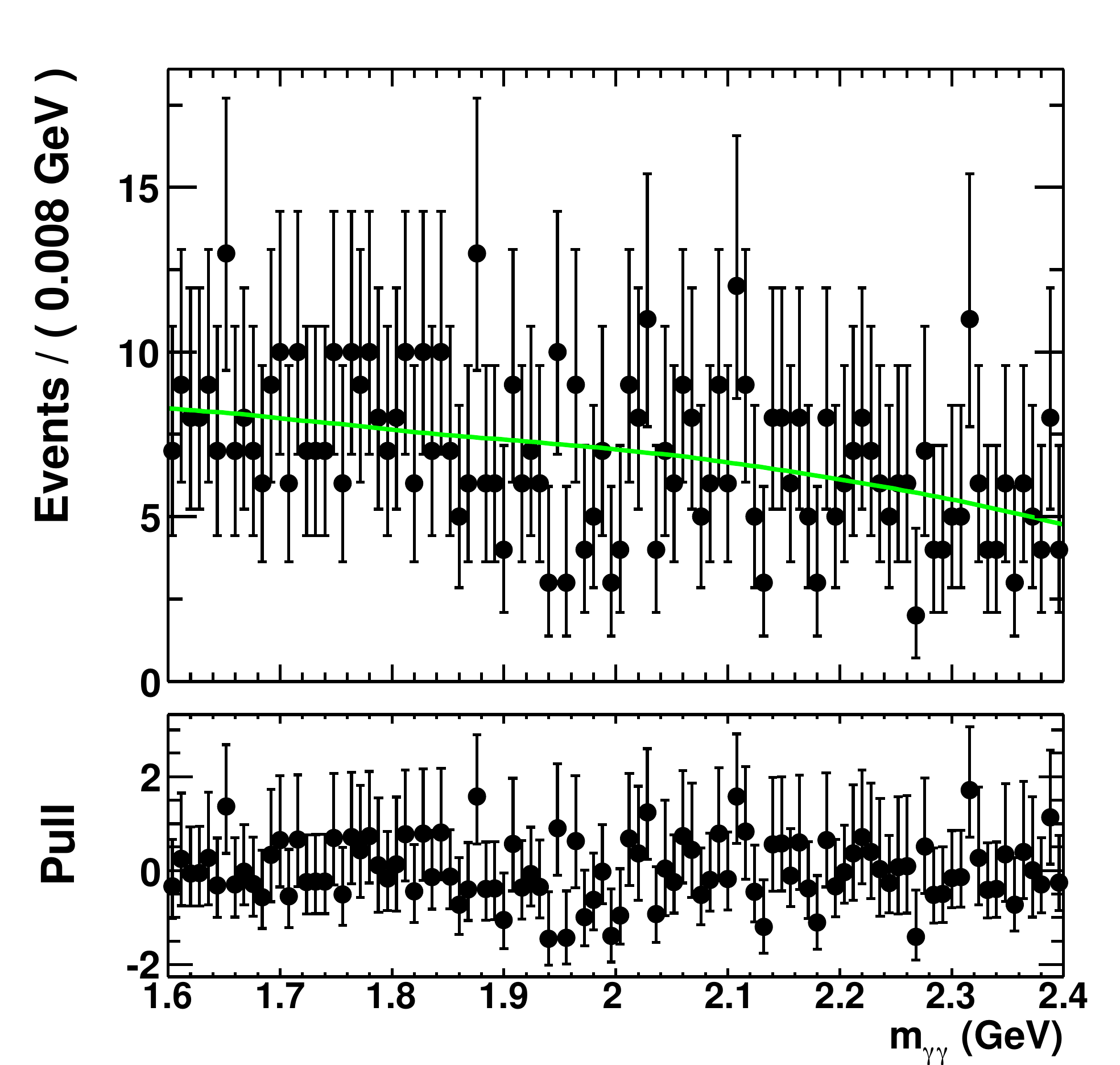}
      \includegraphics[width=0.3\textwidth]{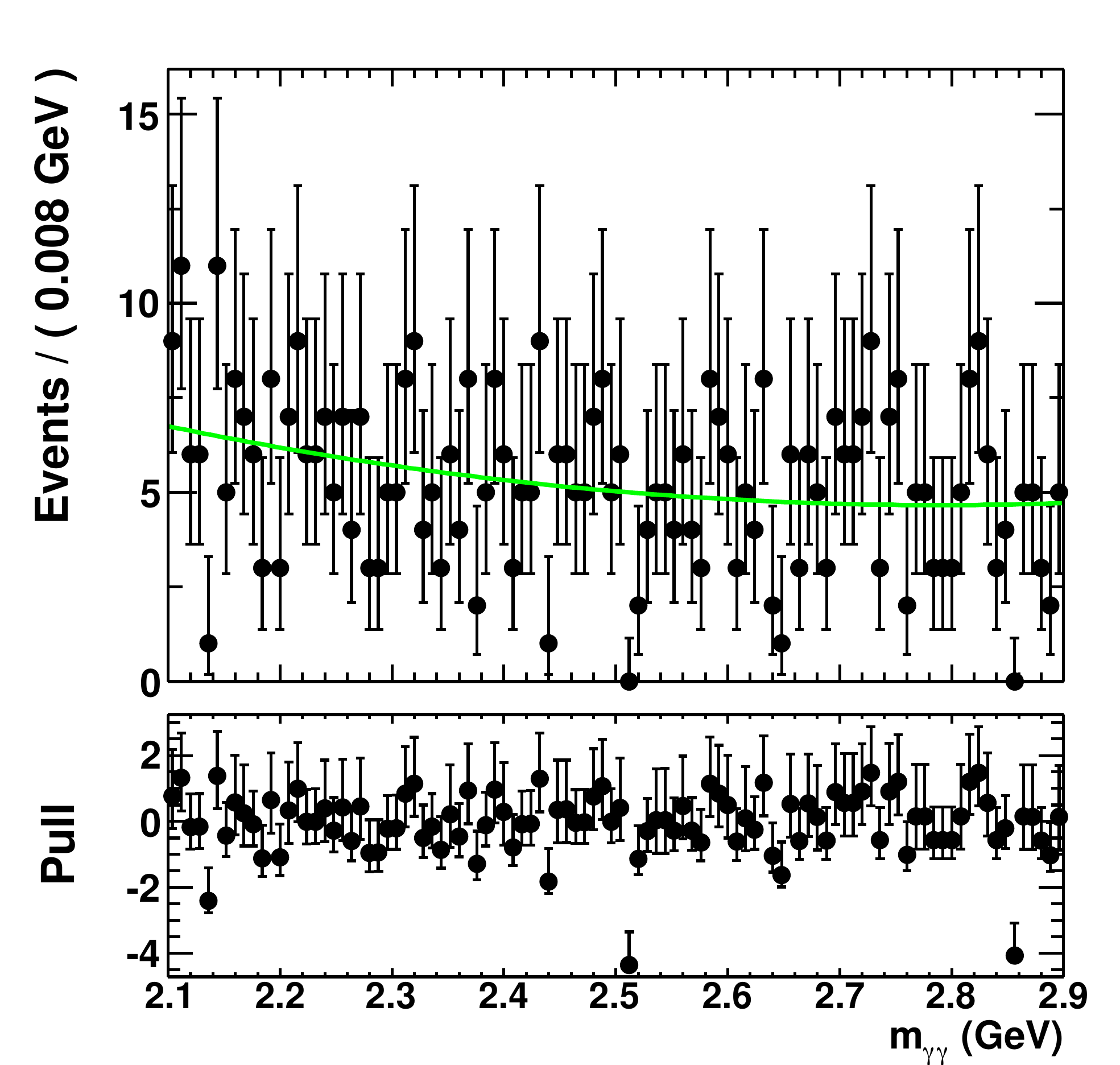}\\
       \includegraphics[width=0.3\textwidth]{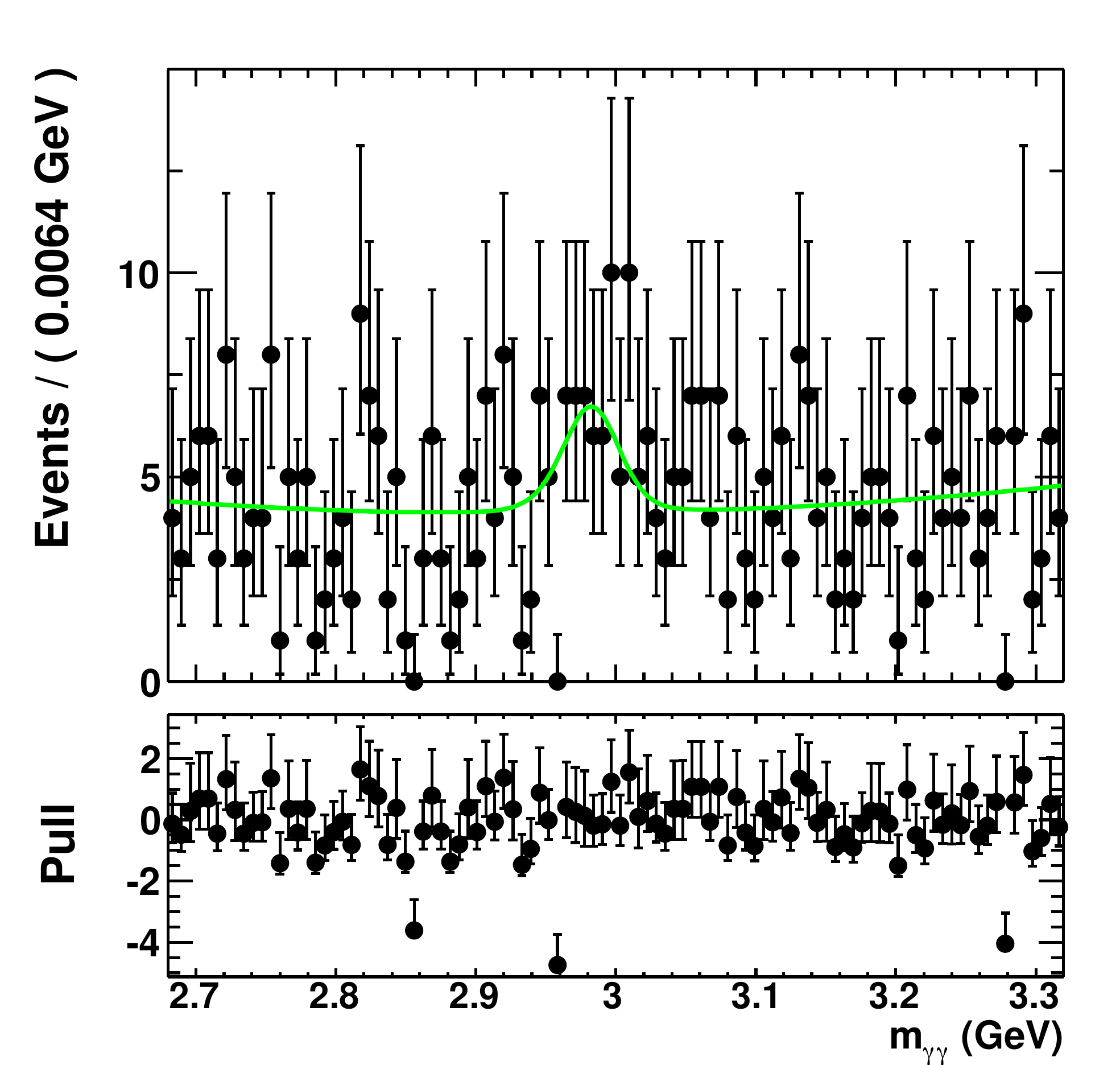}
    \includegraphics[width=0.3\textwidth]{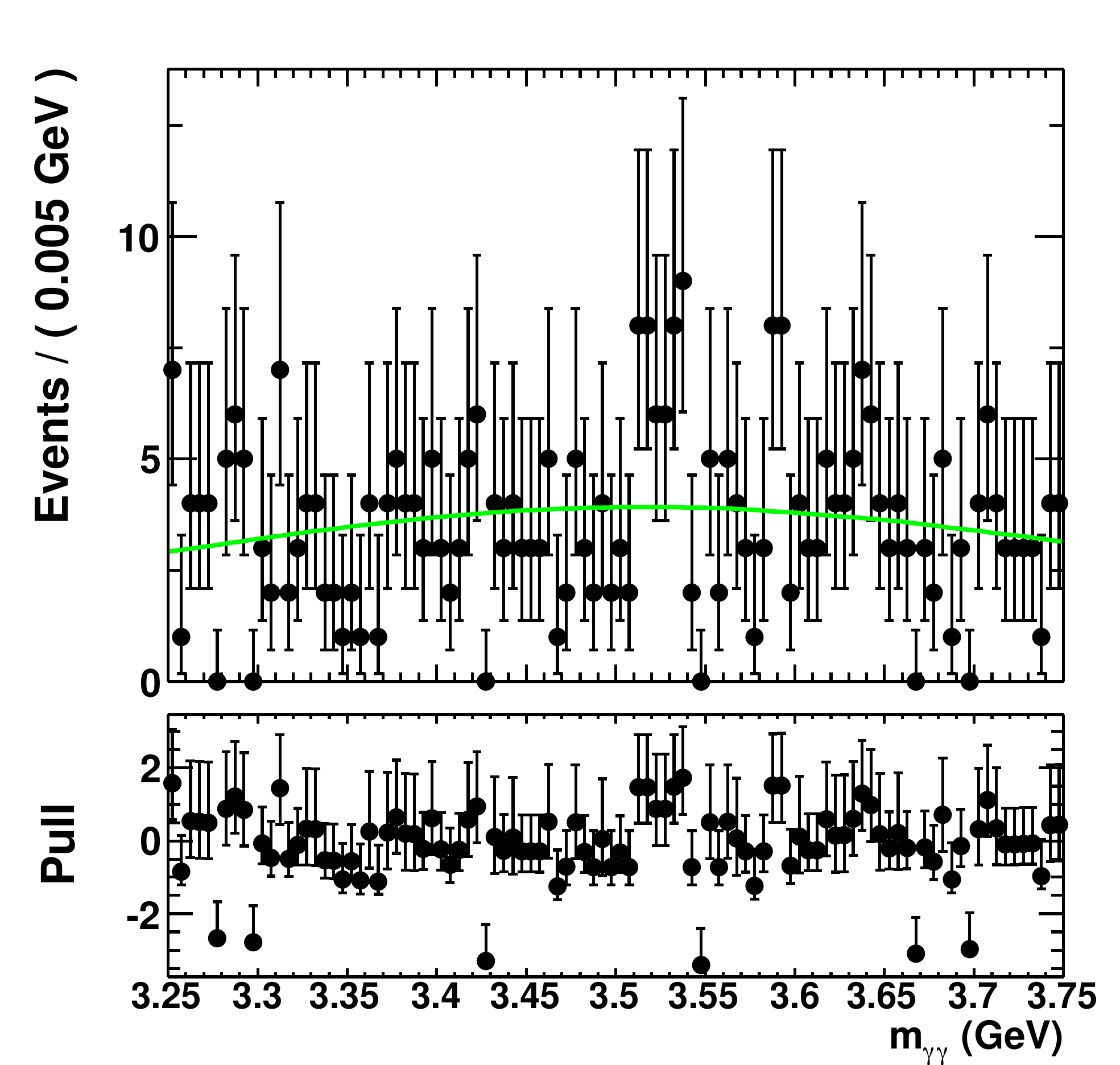}
      \includegraphics[width=0.3\textwidth]{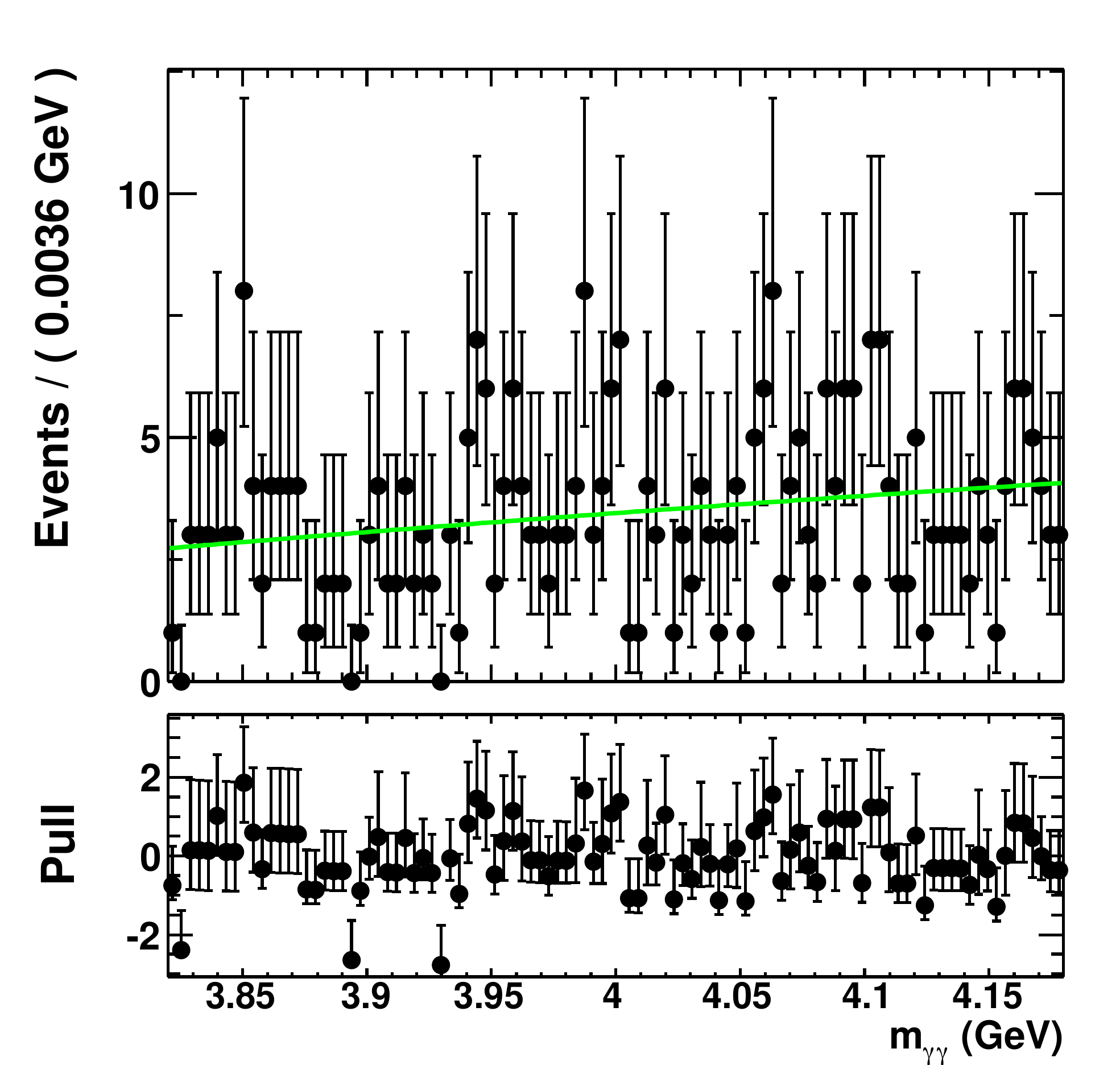}\\
        \includegraphics[width=0.3\textwidth]{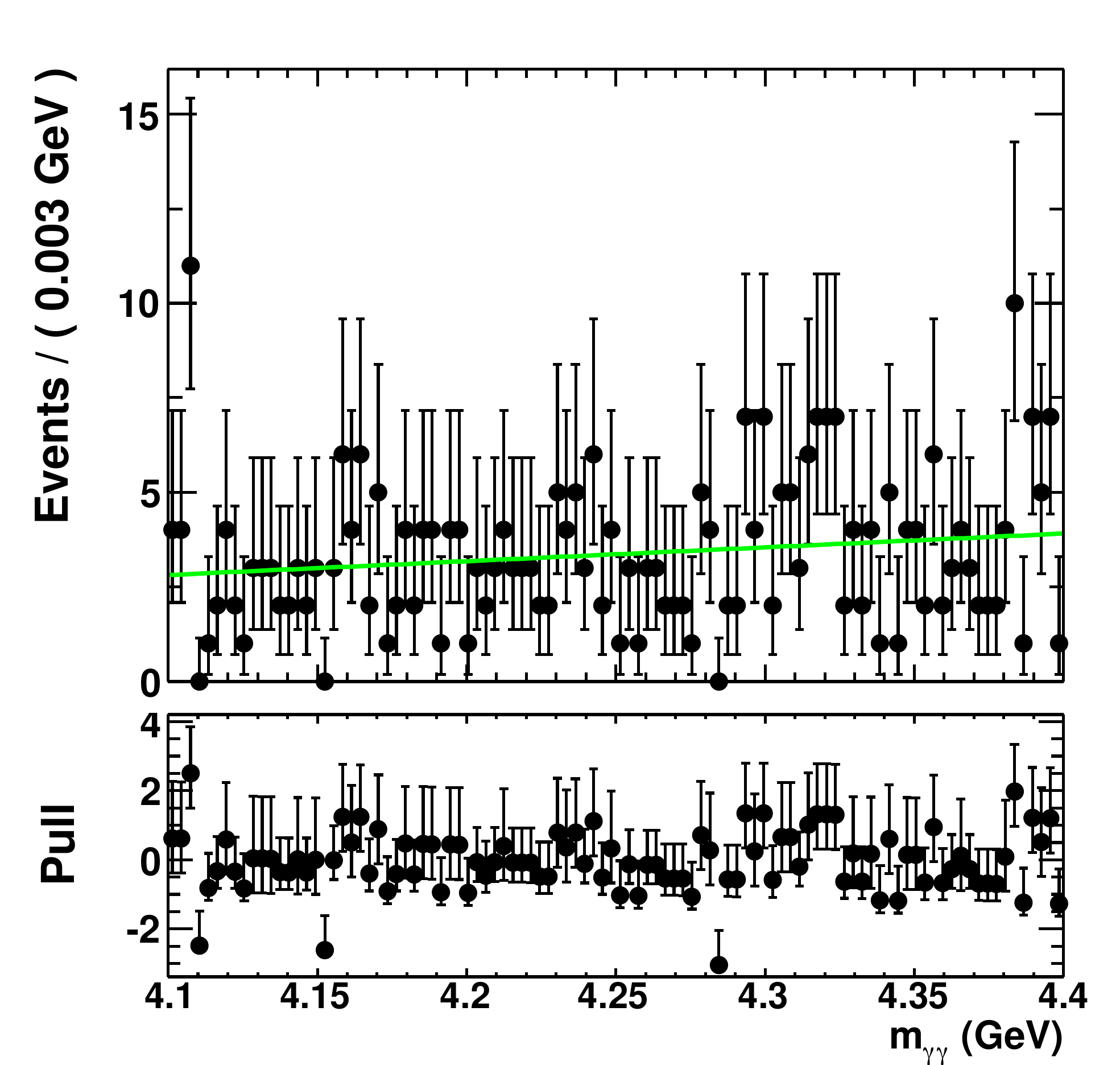}
    \includegraphics[width=0.3\textwidth]{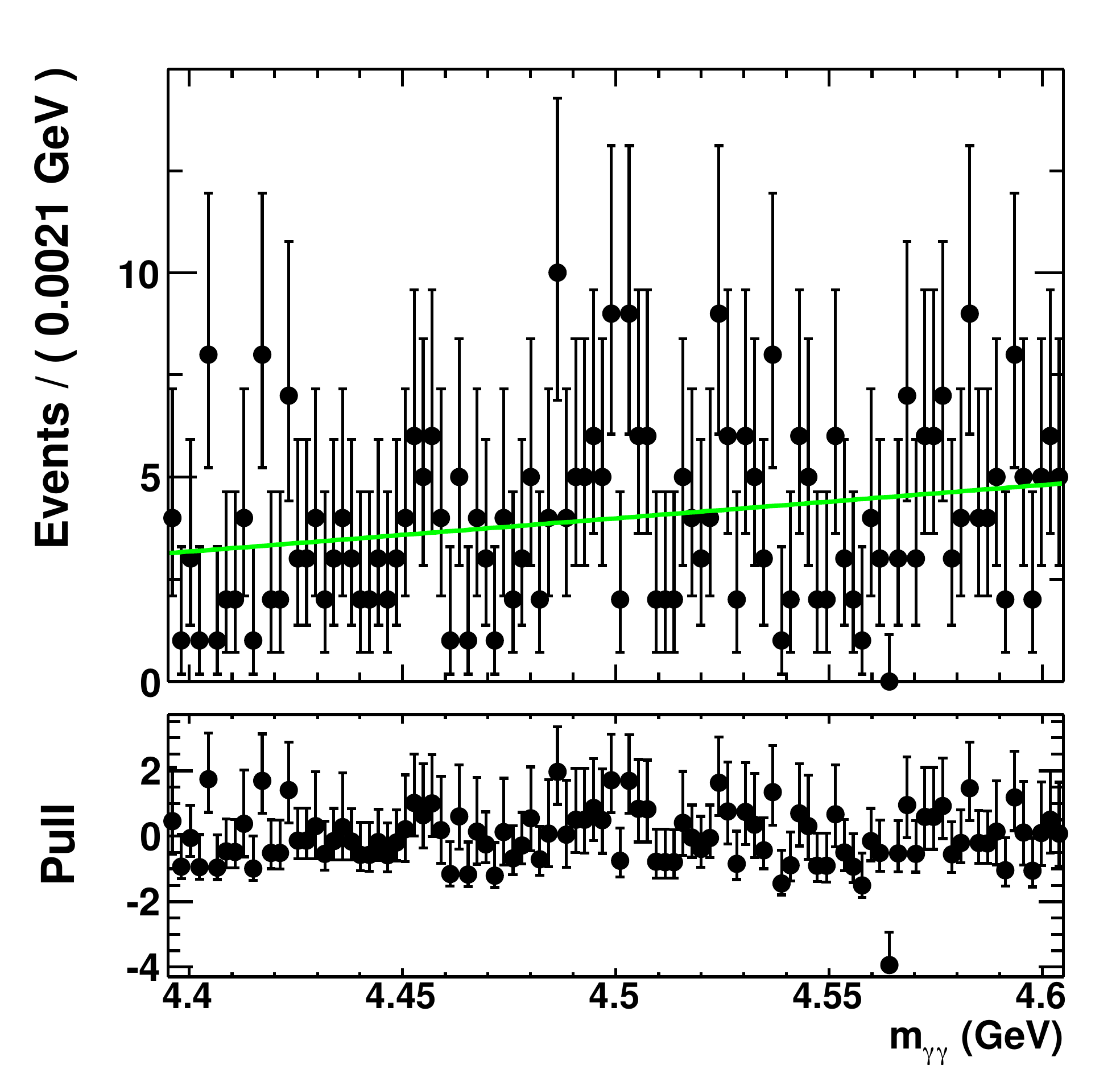}
      \includegraphics[width=0.3\textwidth]{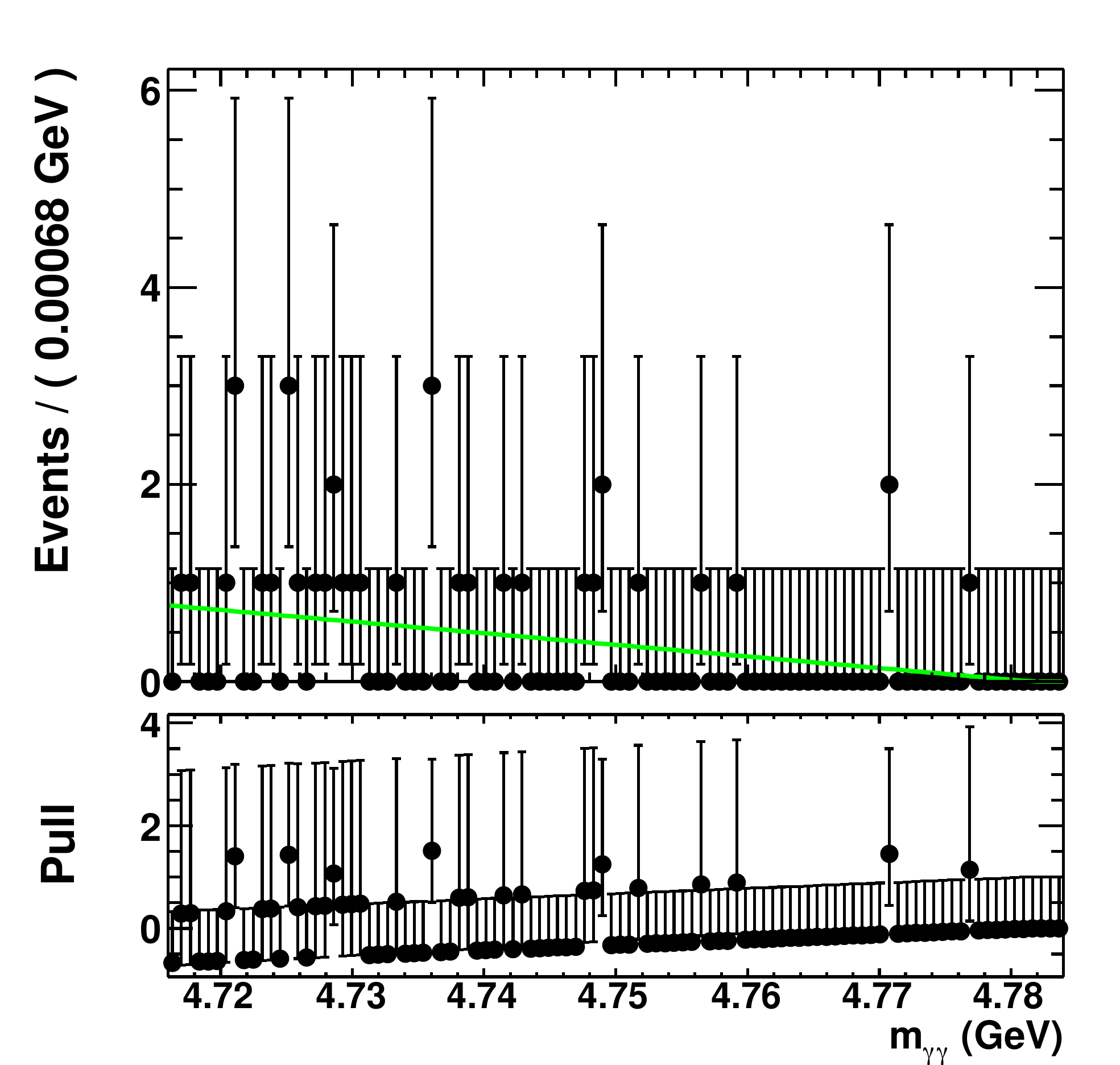}
      \end{center}
      \caption{Fits of our background model to data for a sample of fit intervals used in the ALP search. The green line indicates the best-fit background model, and the points represent data. The fit interval width is determined by the signal resolution as outlined in the main-body text.}
\label{epaps32}
\end{figure}

\begin{figure}[h]
\begin{center}
  \includegraphics[width=0.43\textwidth]{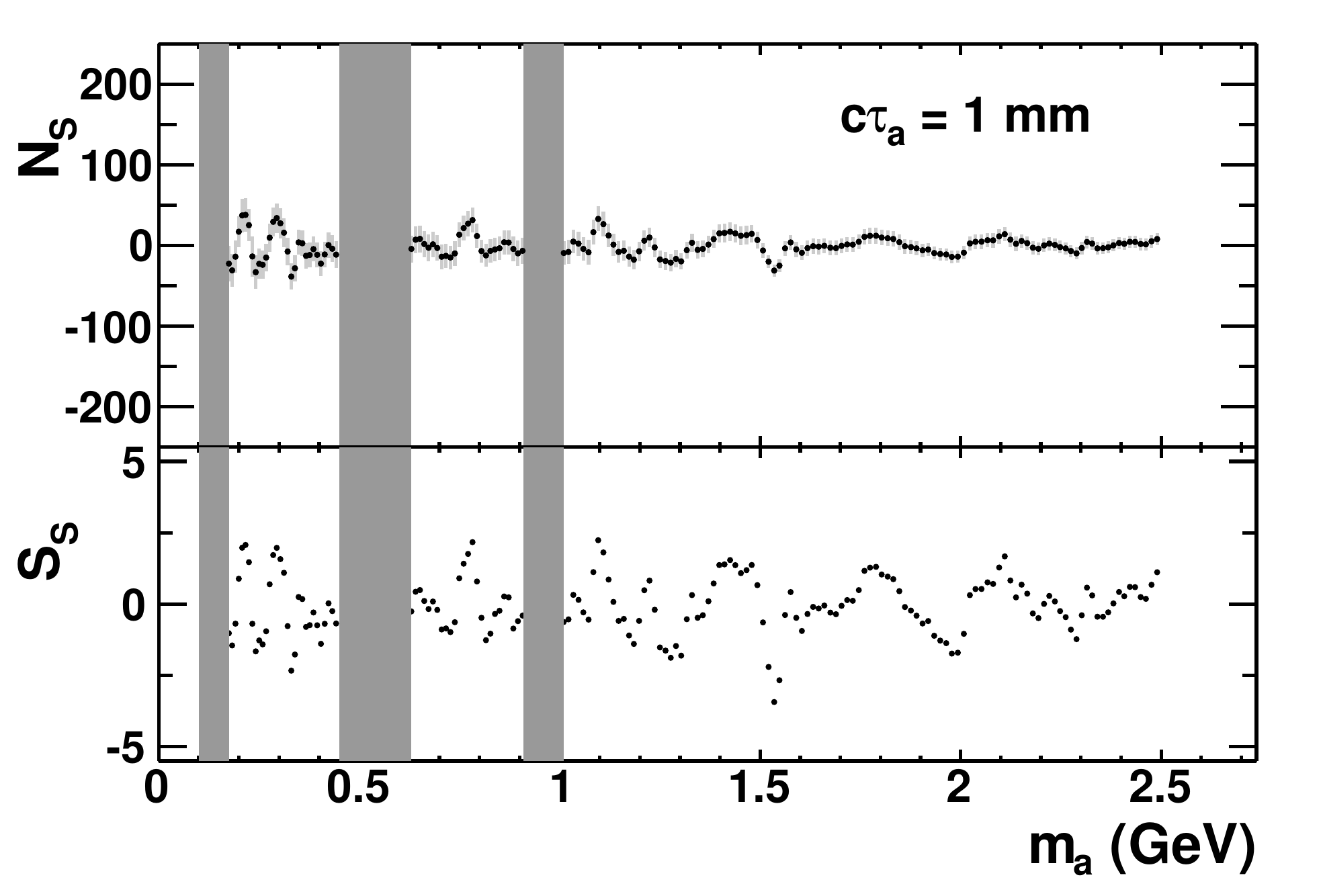}\hspace{1cm}
    \includegraphics[width=0.43\textwidth]{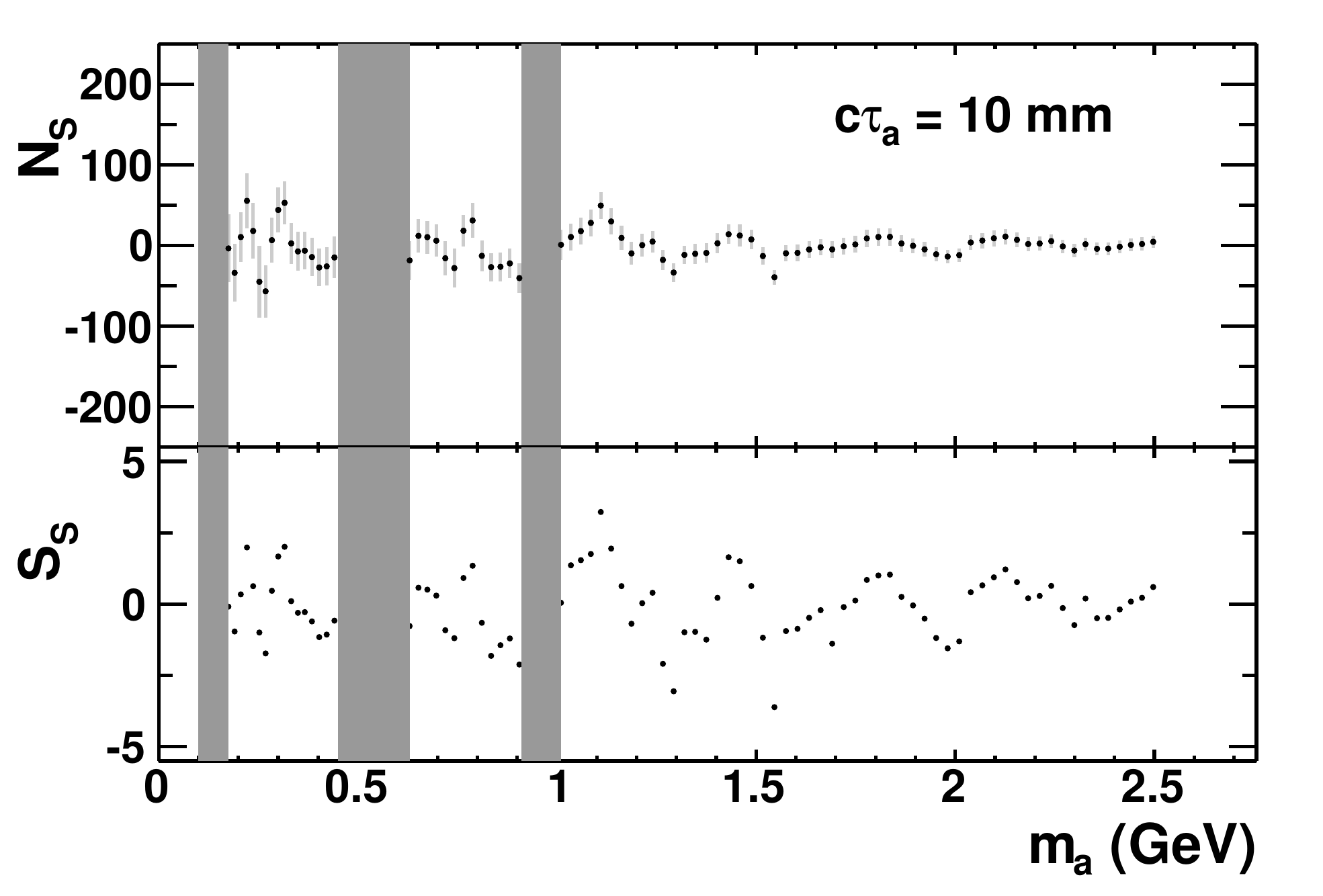} \\
        \includegraphics[width=0.43\textwidth]{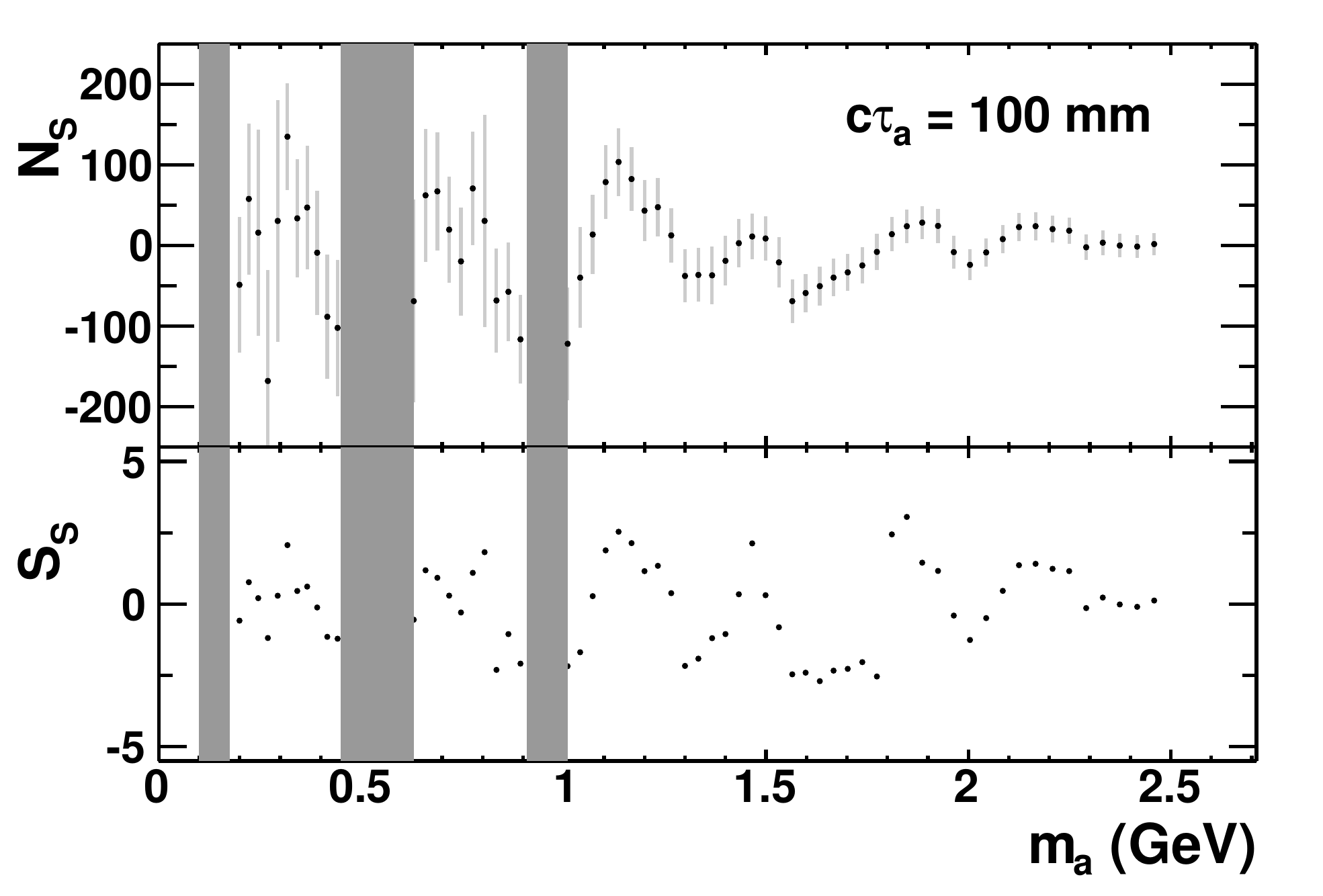} 
\end{center}
\caption{The distribution of signal events ($N_{\rm s}$) and local signal significance ($S_{\rm s}$) from fits as a function 
of $m_a$ for ALP proper decay lengths of 1 mm, 10 mm, and 100 mm. The 
vertical gray bands indicate the regions excluded from the search in the vicinity of the $\pi^0$, $\eta$, and $\eta'$ masses.}
\label{epaps2}
\end{figure}


\begin{thebibliography}{9}
\bibitem{units} Throughout, we use natural units with $c=\hbar=1$ for all energies and  masses.
\end{thebibliography}
\end{document}